\documentclass[10pt,conference]{IEEEtran}
\usepackage{cite}
\usepackage{amssymb,amsfonts}
\usepackage{algorithmic}
\usepackage{graphicx}
\usepackage{textcomp}
\usepackage{xcolor}

\usepackage{array}
\usepackage{color} 
\usepackage{listings}
\usepackage[cmex10]{amsmath}
\usepackage{url}

\usepackage{breakurl}
\usepackage{todonotes}
\usepackage{setspace}
\usepackage{alltt}
\usepackage{hyperref}
\usepackage{framed}
\usepackage{tabu}
\usepackage{subcaption}

\newcommand\david[1]{{}}

\newcommand\lori[1]{{}}
\newcommand\tamara[1]{{}}

\usepackage{csvsimple,booktabs,multirow,makecell}
\def\myrot#1{\rotatebox{90}{\csname csvcol#1\endcsname\ }}

\definecolor{darkgreen}{RGB}{0,128,0}
\definecolor{darkblue}{RGB}{0,0,128}
\definecolor{darkred}{RGB}{192,0,0}

\def\ttob{\texttt{\symbol{`\{}}}
\def\ttcb{\texttt{\symbol{`\}}}}
\newcommand{\ttc}[1]{\ttob{#1}\ttcb}

\def\ttt{\texttt}

\def\trm{\textrm}
\def\tbf{\textbf}

\newcommand{\itc}[1]{\textit{\it #1\/}}

\def\<{\langle}
\def\>{\rangle}

\input{mathmode-spacing.tex}

\def\BibTeX{{\rm B\kern-.05em{\sc i\kern-.025em b}\kern-.08em
    T\kern-.1667em\lower.7ex\hbox{E}\kern-.125emX}}
\begin{document}

\title{Automated Code Repair for C/C++ Static Analysis Alerts}


\author{\IEEEauthorblockN{David Svoboda, Lori Flynn, William Klieber, Michael Duggan, Nicholas Reimer, Joseph Sible}
\IEEEauthorblockA{Software Engineering Institute\\
Carnegie Mellon University\\
Pittsburgh, PA\\
\ttt{\{svoboda,lflynn,weklieber,mwd,nreimer,jcsible\}@sei.cmu.edu}\\}
}

\maketitle
\pagestyle{plain}

\begin{abstract}
{\em Note: This work is a preprint.} Static analysis (SA) tools produce many diagnostic alerts indicating that source code in C or C++ may be defective and potentially vulnerable to security exploits. Many of these alerts are false positives. Identifying the true-positive alerts and repairing the defects in the associated code are huge efforts that automated program repair (APR) tools can help with. Our experience showed us that APR can reduce the number of SA alerts significantly and reduce the manual effort of analysts to review code.
This engineering experience paper details the application of design, development, and performance testing to an APR tool we built that repairs C/C++ code associated with $3$ categories of alerts produced by multiple SA tools. Its repairs are simple and local.
Furthermore, our findings convinced the maintainers of the CERT Coding Standards to re-assess and update the metrics used to assess when violations of guidelines are detectable or repairable.
We discuss engineering design choices made to support goals of trustworthiness and acceptability to developers.
Our APR tool repaired 8718 out of 9234 alerts produced by one SA tool on one codebase. It can repair 3 flaw categories.
For 2 flaw categories, 2 SA tools, and 2 codebases, our tool repaired or dismissed as false positives over 80\% of alerts, on average. Tests showed repairs did not appreciably degrade the performance of the code or cause new alerts to appear (with the possible exception of \ttt{sqlite3.c}).
This paper describes unique contributions that include a new empirical analysis of SA data, our selection method for flaw categories to repair, publication of our APR tool, and a dataset of SA alerts from open-source SA tools run on open-source codebases. It discusses positive and negative results and lessons learned.
\end{abstract}
\IEEEpeerreviewmaketitle

\IEEEpeerreviewmaketitle

\section{Introduction}

\newcommand{\textfrac}[2]{\frac{\textrm{#1}\vphantom{X^{X}}}{\textrm{#2}\vphantom{X_{X_g}}}}

{\em Note: This work is a preprint.}
Static analysis (SA) is a standard method of inspecting code for flaws that could lead to vulnerabilities and misfunction.
Manual adjudication of SA alerts as true or false positive requires costly expertise and time, and often there are too many alerts.
Consequently, some alerts are prioritized for adjudication, leaving others uninspected, resulting in unknown risks in production code.

In this paper, we present our engineering experience developing
an automated program repair (APR) tool called Redemption to help developers handle large sets of alerts in C or C++ code.
The tool accomplishes this by repairing as many alerts as it can in $3$ flaw categories: null pointer dereferences, uninitialized value reads, and code that has no effect. We chose these categories because they produced a large number of alerts, are critical to security, and are well-known. Section~\ref{section.repair.category.analysis} details our selection process.

False positives are a major problem of SA tools, as no tool can correctly adjudicate all alerts with 100\% accuracy. We must therefore assume that some of the alerts to be repaired might be false positives and that we might be unable to judge whether they are false positives. We therefore seek to repair the code anyway. If our APR tool can reliably determine that the alert is a false positive, our tool can report the alert as false positive rather than repairing it. In general, developers prefer to identify and discard false positives rather than repair them; our strategy to repair false positives presumes that they cannot be reliably identified.

Designing an APR tool that preserves good code behavior introduced many engineering considerations, which we discuss in a future section. For example, a good repair conforms to the program's strategy of error handling, but determining this strategy can be tricky. Yet, we were able to accurately repair $94\%$ of the alerts in one codebase, and we showed that APR can eliminate a large percentage of SA alerts, thereby reducing the manual effort of analysts when reviewing code.

The success of our APR tool prompted an update of the metrics in the CERT Coding Standards, as detailed in Section~\ref{subsection:remediation.cost}.

This paper is organized as follows. Section~\ref{section.problem} describes the magnitude of the problem.
Section~\ref{section.repair.category.requirements} analyzes the requirements for a flaw category to qualify for APR.
Section~\ref{section.repair.category.analysis} analyzes the CERT C guidelines \cite{cert.standards} based on the requirements in the previous section and identifies the $3$ guidelines selected.
Section~\ref{section.engineering.challenges} discusses obstacles and difficulties we encountered when building the tool, and Section~\ref{section:case.study} describes the results of testing our solution to determine if it fulfilled our requirements.
Section~\ref{section.related.work} discusses related work, with other APR strategies and solutions.
Section~\ref{section.conclusion} concludes, summarizing positive and negative results and lessons learned.

This paper's contributions include the following: Our case study analyzing and targeting code flaws to automatically repair, specific code/algorithm examples of repairs, details of our engineering implementation, and discussion of issues to consider when creating a system to do automated code repair. We also published the code for our APR tool~\cite{tool.redemption}
and a dataset~\cite{svoboda2023frequency.dataset}
to accompany this paper so that our results can be validated.

\section{The Problem}
\label{section.problem}

The effort to audit and repair the average C/C++ codebase should be measured in {\bf{person-years}}. In a NIST study, $66\%$ of application security tool findings were determined to be irrelevant~\cite{delaitre2018sate}. Another study found that triaging each finding takes an average of $10$ minutes~\cite{anderson2012measuring}. According to CodeDx~\cite{synopsys.codedx},
``A tool returns an average of $10,000$ results on an application. With a $66\%$ irrelevancy
rate, this is equal to $132$ days spent reviewing false positives.''

We also analyzed audit and repair effort. The analysis is based on defect density, which is the ratio of the total number of defects in a codebase to its size, which we measured in thousand lines of significant code (kSigLoC). The defect density of the C/C++ codebases in our analysis is \,$85{,}268$ alerts / \,$233.9$ kSigLoC \,$= 364.5 \,\,\textfrac{alerts}{kSigLoC}$. Suppose that the average C/C++ codebase is $1{,}957$ kSigLoC per a set of proprietary codebases, which includes the ones in our analysis,
and the average kSigLoC produces $364{.}5$ SA alerts.
According to a large study by Google, it took an average of $117$ seconds to audit one alert~\cite{ayewah2010google}
and let us guess another $117$ seconds is required to fix each alert. 
(Some alerts, like buffer overflows, usually take much more than $117$ seconds to fix, but we focus on alerts that are simple to fix, such as null pointer dereferences. If a developer makes many simple fixes at once, then $117$ seconds per alert becomes reasonable.)
The developers in the study manually fixed
$32\%$ of the alerts~\cite{ayewah2010google}.
So, the average time for a person to deal with an alert (i.e., to audit it and fix it if necessary) was
$
117\textrm{ sec} + 32\% * 117\textrm{ sec} = 154.44\textrm{ sec}
$.
Using the above-mentioned defect density of 364.5 alerts per kSigLoC results in an average per-person effort (for a kSigLoC) of $154.44 \textfrac{sec}{alert} * 364.5 \textfrac{alerts}{kSigLoC} = 56{,}293 \textfrac{sec}{kSigLoC}$.
Thus, implied effort for an average codebase of 1,957 kSigLoC is
\vspace{-2ex}
$$1{,}957\trm{ kSigLoC} * 56{,}293 \textfrac{person$\cdot$sec}{kSigLoC}
= 3.5\trm{ person$\cdot$years}.$$
\vspace{-2ex}

 Most organizations do not have this much effort available to address SA alerts, and they consequently ignore most alerts, leaving many potential true positives unrepaired.

\section{Repair Category Requirements}
\label{section.repair.category.requirements}

To qualify as repairable, a code defect must have a repair solution that can be automatically implemented by a repair tool. The tool may use the \emph{abstract syntax tree} (AST) or \emph{intermediate representation} (IR) of the code in which the repair will be placed. What follows are several attributes to a guideline and their suitability for our APR tool.

\tbf{Trustworthy:}
Repairs should not ``break'' the code, whether the alert is true or false. This means APR patches should not change functionality; add security flaws; cause unit, performance, or integration tests to fail; or disable compilation.

For example, the typical repair for an alert about a memory leak is to add a call to \ttt{free()}. However, if the code is correct initially and the alert is a false positive, adding a call to \ttt{free()} is likely to cause either a memory leak or a double-free. Consequently, repairs for memory leaks are ``breaking'' repairs. See Section~\ref{subsection:trustworthy.repairs} for more information.

\tbf{Locality:}
While some vulnerabilities can be assigned to a specific line of code, they require changes in multiple places in the codebase. They can't be fixed by changing just the line of code or just the function that contains the code.
For example, buffer-overflow alerts are typically not locally repairable if the buffer was allocated in a different function than that of the alerted line of code because there is no way to determine the capacity of the buffer using only information available at the alerted line of code.

\tbf{Developer Acceptance:}
Once a repair is applied, it has two liabilities. First, it may alter the intended behavior of the code or even break it. Second, the repair might preserve intended behavior but be rejected by developers in favor of a manual repair or even no repair to the code. Determining what changes developers will accept is complicated, and we do not fully address it here. We believe developers are less likely to accept repairs that change many lines of code than repairs that change one line or a few lines of code~\cite{sadowski2015tricorder,mechtaev2015directfix}.

For example, it is better to reliably determine that an alert is a false positive rather than to repair it, as making any change to source code requires developers to learn the change. As another example, a repair tool should not repair an alert that has already been repaired, as a double repair may not change how the code behaves but would be rejected by developers.

\tbf{Error Handling:}
Some guidelines can be repaired but require an error-handling strategy if the repair can determine that the code is about to violate the guideline. Which error-handling strategy to use depends on the codebase's error-handling policy and may also depend on the particular function. As an engineering choice for our tool, we assume the existence of a ``standard'' error-handling mechanism. Our error-handling strategy is expected to prevent execution from continuing, which prevents the defect from happening.

For example, a null pointer dereference can be repaired by inserting (before the dereference) comparison of the pointer with null and then handling the error. This repair is local. A null pointer dereference repair for a false-positive alert means that the code receives an extra null check on a pointer that can't be null. This introduces a minuscule slowdown on the code. Null-dereference repairs are non-breaking. For more information, see Section~\ref{subsection:error.handling}.

\tbf{Static Analysis:}
Static analysis is usually best done on an intermediate representation such as LLVM IR. However, repair is usually easiest to do using the AST,
since mapping AST-level changes back to the source-code text is easier than mapping IR-level changes. For the most part, our repair tool does not perform its own static analysis; instead, it relies on the output of external SA tools. While some SA tools put great effort into preventing false positives, not all SA tools do. A potential problem is that an SA tool might continue to flag a piece of code after our tool has applied a repair to it. This is undesirable because applying our tool again in the future will add redundant repairs to the code. So, our repair tool does a small amount of its own static analysis to avoid repairing an alert that has previously been repaired. Also, if our tool detects that an alert is \itc{dependent}~\cite{svoboda2016static} on another alert that is being repaired, it does not repair the dependent alert.

\tbf{Automatic Detection:}
This turns out not to be required for repairable categories. A flaw category might be automatically repairable while not being automatically detectable. For example, null pointer dereferences can be automatically repaired, but accurate automatic detection of null pointer dereferences is a hard problem.

\section{Repair Category Analysis}
\label{section.repair.category.analysis}

Our tool is designed to repair alerts that report flaws. Flaws can be partitioned by various coding standards, such as the CERT C Coding Standard~\cite{cert.standards} or MITRE CWEs~\cite{mitre:cwe}. This section describes how we selected $3$ CERT C guidelines to repair out of its more than $300$ total guidelines.

\subsection{Remediation Cost}
\label{subsection:remediation.cost}

Each CERT Coding Standard guideline includes several metrics about the cost of complying with the rule. Previously, it employed a ``Remediation Cost'' metric that indicated whether compliance with the rule was automatically detectable or whether an alert was automatically repairable. These metrics presumed that a guideline that can be automatically repaired is not easy to fix unless it is also automatically detectable. As noted in Section~\ref{section.repair.category.requirements}, our approach to APR violates this assumption. The success of this project and tool prompted an adjustment to the metrics. Now, the ``Detectable'' metric addresses whether a static analysis tool can automatically determine if code violates the guideline with high precision and recall. The ``Repairable'' metric addresses whether an automated repair tool can reliably fix an alert by making local changes, plus whether the repair can be guaranteed not to break the code even if the alert is a false positive.

\subsection{CERT Guideline Impact}
\label{subsection:cert.rule.impact}

The simplest metric to rate guidelines' impact is to count how many alerts map to each guideline in a set of SA alerts for a set of codebases. In a preliminary analysis, we used summary SA statistics from C/C++ codebases that was performed as part of SCALe audits \cite{project.scale} \cite{training.scale}. These codebases consisted of $234$ kSigLoC (where a kSigLoC is $1,000$ lines of code excluding blank lines and comments). These codebases produced $85{,}268$ alerts that allegedly violated $124$ distinct CERT C or C++ guidelines \cite{cert.standards}. Fortunately, $57{,}922$ alerts (67.9\%) violated just 8 CERT guidelines, as shown in Table~\ref{8.cert.rules.latex}. If we automatically repaired 80\% of these 8 guideline-violations, that would fix 54.3\% of the alerts.
(When we speak of repairing an alert, we are actually repairing the defect in a codebase that causes the alert to be reported.)  The first rule alone (INT31-C) generates $21{,}264$ alerts or 24.9\% of the alerts! Fixing 80\% of these resolves 20.0\% of all alerts. A similar analysis on different codebases might produce different “top 8” CERT guidelines.

\begin{table}[htbp]
\caption{8 CERT Guidelines: Title, Pri(ority), Alert Count}
\vspace{-1ex}
		\begin{center}
			\begin{tabular}{|>{\raggedright\arraybackslash}p{1.5cm}
					|>{\raggedright\arraybackslash}p{4cm}
					|>{\raggedleft\arraybackslash}p{.6cm}
					|>{\raggedleft\arraybackslash}p{1cm}|}
				\hline
\bf{Guideline} & \bf{Title} & \bf{Pri} & \bf{Alerts}\\
				\hline
				INT31-C & Ensure that integer conversions do not result in lost or misinterpreted data & 6 & $21,264$ \\
				\hline
				INT13-C & Use bitwise operators only on unsigned operands & 6 & 11,007 \\
				\hline
				EXP00-C & Use parentheses for precedence of operation & 4 & 6,602 \\
				\hline
				PRE03-C & Prefer \ttt{typedef}s to defines for encoding types & 2 & 6,093 \\
				\hline
				DLC00-C & Const-qualify immutable objects & 1 & 4,863 \\
				\hline
				EXP05-CPP\footnotemark & Do not use C-style casts & 12 & 3,294 \\
				\hline
				EXP12-C & Do not ignore values returned by functions & 4 & 2,905 \\
				\hline
				MSC13-C & Detect and remove unused values & 2 & 1,894 \\
				\hline
			\end{tabular}
			\label{8.cert.rules.latex}
					\end{center}
                \vspace{-3ex}
\end{table}

\footnotetext{EXP05-CPP has been withdrawn, but it appears in older SCALe audits.}

\subsection{Flaw Frequency Background}
\label{subsection:flaw.frequency.background}

Determining which code defects or alerts are most common requires a big dataset, and the choice of dataset heavily influences which alerts are most common. Many codebases are not public, nor are their SA results (whether or not they include manual adjudications of the alerts as true or false positive). Theoretically, the study of large datasets like GitHub code scans could be helpful. GitHub requires that users have write permission to view an alert summary for a repository~\cite{github-codescanning}.

One study involving Coverity SA tool's alerts for four large, highly-used, open-source codebases (each over $100{,}000$ lines of code) found only 67 of the possible 193 alert types were produced, and 80\% of the alerts came from ~20\% of the alert types. They published the top 10 types and average time until manual code fixes for actionable alerts per type ($24-660$ days)~\cite{imtiaz2019developers}. The top-10 list could be used to target development of APR for code flaws that could cause those alerts, and APR time-savings could be estimated using their fix-time data. Our selected CWE-476 ``NULL pointer dereference'' relates to items on their list: ``4. Explicit null dereferenced'', ``5. Dereference after null check'', ``6. Dereference before null check'', and ``8. Dereference null return value''. Our selected CWE-561 ``Dead Code'' relates to their item ``3. Logically dead code''. Our selected CWE-908 ``Use of Uninitialized Resource'' relates to (but entails more than) their item ``9. Uninitialized scalar variable''. A large-scale analysis of C/C++ vulnerabilities in the CVE database and associated open-source code from 2006 to 2023 listed the top 10 CWEs associated with the CVEs~\cite{ni2024megavul}. Our selected CWE-476 is 4th on their list, but CWE-908 and CWE-561 are not on it.

Some SA tool vendors publish aggregated results of scans with their tools that include flaw frequency per language. For example,  Veracode reports are currently based on assessments of over 14 trillion lines of code, though the flaw types listed are not as granular as CWEs~\cite{veracode2020state}.

A codebase or development team may have an unusual set of most-frequent code defects, due to their coders' strengths and weakness, tools, and the codebase. Analysis of their alerts could help target code repairs to develop for them.

Alert counts or true-positive alert counts are not the only factors to consider when prioritizing effort to develop automated repairs. Other important factors are the estimated impact of the code defect if not fixed, amount of human effort required to make the automated fix (e.g., if a human needs to provide input to guide the particular automated repair used), and impact of a repair on code comprehension by a developer that affects code maintenance.

\subsection{Frequency Analysis}
\label{subsection:frequency.analysis}
Based on the preliminary analysis referenced by Section~\ref{subsection:cert.rule.impact}, we conducted a frequency analysis of SA alerts produced using open-source SA tools on several open-source codebases. The SA tools we used are
clang-tidy version 15.0.7~\cite{tool.clang.tidy},
Cppcheck version 2.9~\cite{tool.cppcheck},
and CERT Rosecheckers~\cite{tool.rosecheckers}.
The codebases we analyzed are
zeek version 5.1.1~\cite{tool.zeek}
and git version 2.39.0~\cite{tool.git}.
We also analyzed dos2unix version 7.4.3~\cite{tool.dos2unix}
but did not include its results here. Our dos2unix data is in our related dataset publication~\cite{svoboda2023frequency.dataset}. Dos2unix is a small codebase requiring the compilation of just $3$ C files. It is too small to produce useful results for publication, but its size works well for development testing.

To measure the frequency of each guideline, we attempted to map each alert to a CERT guideline. Some alerts had no suitable CERT guideline to map to. A small sample of our results are shown in Table~\ref{table:rules.data.tools.codebases}. Each row lists a CERT guideline, and the columns indicate how many alerts that map to the guideline were reported by the corresponding tool on the corresponding codebase. For each tool and codebase combination, there is a ``Rank'' column; this is used to indicate the most prolific guidelines with flaws reported by the tool on the codebase. The two CERT rules also list their priority, plus an assessment of whether the rule is automatically repairable or not. The table concludes with a ``Total'' row, which contains the sum of all alerts that map to a CERT guideline, and a ``Guidelines'' row, which contains the number of distinct CERT guidelines for which there was at least one alert.

So that others can validate our analysis, we published data we produced and details about the open-source tools and code we used~\cite{svoboda2023frequency.dataset}. (Note: this dataset may differ slightly from the numbers reported in this paper due to slightly different versions of the tools used.)
The dataset includes instructions for running the SA tools, a Dockerfile to conveniently obtain the SA tools, raw SA tool output, parsed SA data and aggregate analyses, and SA data augmented with CERT guideline and CWE data.

\begin{table*}[htbp]
  \small
  \caption{Excerpt of alert counts and ranking for tools, codebases, and guidelines. ``Total'' is the sum of alerts for the tool and codebase, and ``Guidelines'' is the total number of CERT guidelines flagged by alerts.}
    \vspace{-1ex}
    \begin{center}
    {
    \renewcommand{\arraystretch}{1.2}
    \begin{tabu}{|l|c|c|*{6}{rr|}}
      \hline
      \multicolumn{3}{|c|}{} & \multicolumn{6}{c|}{git} & \multicolumn{6}{c|}{zeek} \\
      \cline{4-15}
      \multicolumn{3}{|c|}{} & \multicolumn{2}{c|}{Cppcheck} & \multicolumn{2}{c|}{clang-tidy} & \multicolumn{2}{c|}{Rosecheckers} & \multicolumn{2}{c|}{Cppcheck} & \multicolumn{2}{c|}{clang-tidy} & \multicolumn{2}{c|}{Rosecheckers} \\
      \hline
      Guideline & Priority & Repairable & Rank & Alerts & Rank & Alerts & Rank & Alerts & Rank & Alerts & Rank & Alerts & Rank & Alerts \\
      \hline

      MSC13-C & 2 & yes & 1 & 228 & 9 & 458 &    &      & 1 & 1179 &  & 16 &   &      \\ \hline
      DCL19-C & 2 & no  & 2 &  63 &   &     &    &      & 2 &  434 &  &    &   &      \\ \hline
      DCL01-C & 2 & yes & 3 &  42 &   &     &  4 & 2465 & 4 &   89 &  &    & 1 & 2553 \\ \hline
      MSC12-C & 2 & yes & 4 &  25 &   &     & 10 &  721 & 3 &  131 &  &    & 8 &  480 \\ \hline
      \hline\hline
      Total         &  &  &  &  420 &  & 48,233 &  & 27,162 &  & 2466 &  & 18,806 &  & 10,885 \\ \hline
      Guidelines    &  &  &  &   13 &  &    29 &  &    49 &  &   24 &  &    41 &  &    45 \\ \hline
    \end{tabu}
    }
  \end{center}
  \label{table:rules.data.tools.codebases}
  \vspace{-2ex}
\end{table*}

\subsection{Analysis Summary and Decisions}

In deciding which guidelines to repair, we considered alert frequency, CERT Prioritization metrics, and guidelines that have analogous CWEs in the CWE Top 25~\cite{mitre:cwe.2021.top.25}.
We used these to rank the CERT guidelines from most to least worthy of our efforts, then selected the following top $3$ for our work: 

\subsubsection{EXP34-C (CWE-476) Null-Pointer Dereferences}
\begin{itemize}
  \item 5th most alerts generated by Cppcheck on git
  \item 6th most alerts generated by Cppcheck on zeek
  \item 11th in the 2022 CWE Top 25 Vulnerabilities
  \item Priority 18 (the highest) in the CERT C Coding Standard
\end{itemize}

\subsubsection{EXP33-C (CWE-457) Read of Uninitialized Value}
\begin{itemize}
  \item 8th most alerts generated by Cppcheck on git
  \item 2nd most alerts generated by clang-tidy on git
  \item 1st most alerts generated by clang-tidy on zeek
  \item Priority 12 (2nd highest) in the CERT C Coding Standard
\end{itemize}

\subsubsection{MSC12-C (CWE-561, CWE-1164) Remove Code that Has No Effect}
\begin{itemize}
  \item 4th most alerts generated by Cppcheck on git
  \item 10th most alerts generated by Rosecheckers on git
  \item 8th most alerts generated by Rosecheckers on zeek
\end{itemize}

\section{Engineering Challenges}
\label{section.engineering.challenges}

This section covers some of the issues we encountered while building our APR tool and our solutions for addressing them.

\subsection{Implemented Repairs}

In this section, we describe the repair algorithms for the categories we selected for repair.

For a null pointer dereference, we use a \ttt{null\_check()} macro to protect any pointer that might be null. We replace any expression $x$ that might represent a null pointer with the value \ttt{null\_check(x)}. This macro is defined in a file \ttt{acr.h}, and it effectively invokes error-handling code if $x$ is null. Here $x$ can be an expression; it need not be a variable. By default, the error-handling code aborts the program via \ttt{abort()}, but you can override it to have more specific behaviors (e.g., \ttt{return NULL} if inside a function that returns \ttt{NULL} on an error).

For an uninitialized value read, we simply go to the variable's declaration and set it to $0$ (or the equivalent of $0$ for the variable's type; e.g., it could be $0.0$ for floats).

For ineffective code (i.e., code that has no effect), we remove the ineffective portion of the statement or expression. One example of ineffective code is an evaluation of some expression, with the result assigned to a variable that is never read (e.g., \ttt{x = foo(...);}). Our repair for that preserves the expression and removes the assignment (e.g., \ttt{(void) foo(...);}).

Casting a function's return value to \ttt{(void)} is a common indication that the function's return value is to be ignored. See Exception 1 of CERT rule
EXP12-C~\cite{cert.standards.EXP12.C} for more information.
This code is ineffective only if the assignment can be proved not to invoke a C++ constructor, destructor, overloaded assignment operator, or overloaded conversion or cast operators, which might have side effects.
Also, Ineffective Code (MSC12-C) is a CERT recommendation (sometimes correct to do), not a rule (always correct to do), and this presented additional complications. Therefore, we added a \ttt{REPAIR\_MSC12} environment variable so the MSC12-C repair is only done if the user explicitly specifies to include it.

\subsection{Error Handling}
\label{subsection:error.handling}

As previously noted, many automated repairs require the existence of an error-handling strategy that the program can employ if it detects that it can prevent an error, such as dereferencing a null pointer. The C language has many techniques for handling errors~\cite{svoboda2016beyond}, but deciding which error-handling strategy is best to employ can be tricky. CERT guideline ERR00-C recommends determining an error-handling strategy, and many coding projects do so. If a coding project has an error-handling convention, ideally our repair tool would recognize and use it. There are several common conventions:

\begin{itemize}
  \item Terminate the program (e.g., call \ttt{abort()} or \ttt{exit()}).
  \item Return an ``invalid'' value, such as \ttt{NULL} or \ttt{EOF}. Or return prematurely, for void functions.
  \item Transfer control elsewhere (e.g., use \ttt{goto} or \ttt{longjmp()} or raise a signal).
\end{itemize}

The error-handling strategy need not terminate the program, but it must not allow execution to resume normally. While our repair tool can employ some intelligence to determine what error-handling strategy to use, it should also allow users to specify the error-handling strategy.

We conducted an informal analysis on dos2unix~\cite{tool.dos2unix}, git~\cite{tool.git},
and zeek~\cite{tool.zeek} to identify their error-handling strategies. dos2unix doesn't seem to have one. Git has an \ttt{error()} function that takes arguments in the style of \ttt{printf()}, reports the error, and returns. Git also has a \ttt{die()} function that calls \ttt{exit()} (but can probably be overridden). Zeek has a \ttt{sec} enum class that indicates various errors, often used as return values (\ttt{sec::none == 0}), plus several enum classes that indicate various non-error states (e.g., \ttt{connection\_state}).

We examined the null-pointer (EXP34-C) alerts in dos2unix, git, and zeek.
Based on this analysis, we settled on the following heuristic for error handling:

\begin{itemize}
\item If a function returns an \ttt{int} (or any integer type that is not an enum), then look for all return statements. Any return statement with a value distinct from the final return statement (at the end of a function) indicates a suitable value to return upon error.
\item A function that returns an enumeration class is harder, unless the enum class provides an obvious error value that we can use. Our tool does not currently handle that.
\item Likewise, if a function returns a pointer type, then look for all return statements. If it returns null anywhere in the function, but a potentially non-null value at the end, then assume that the function can return null upon error.
\item If the above fails, default to \ttt{abort()}. Custom error-handling code is currently beyond our ability to detect.
\end{itemize}

Our tool allows the user to provide custom error-handling code, such as a \ttt{die()} function.

\subsection{Conditional Compilation Directives}
\label{subsection:conditional.compilation.directives}
C's preprocessor may be the biggest reason why automated repair of C/C++ remains difficult~\cite{medeiros2017discipline}.
The nature of the preprocessor allows the compiler to produce different code depending on which configuration of macro definitions is known to the compiler. Therefore, a C source file represents not one program but a family of one or more programs, each of which differs by a unique configuration of macro definitions.

We decided that our repair tool (like many SA tools and all compilers) will only operate on one macro configuration at a time, and that macro configuration must produce a working C program. Our tool will also strive to ensure that repairs made to one macro configuration will not break other macro configurations. To repair multiple configurations, you must be able to build each such configuration.

One concern with conditional preprocessor directives (e.g., \ttt{\#if} / \ttt{\#ifdef} / \ttt{\#ifndef}) is that not all macro configurations are necessarily compilable. While it is possible to write conditional preprocessor directives such that all configurations compile, this is often not done. If a program fails to compile when the macro \ttt{FOO} is not defined, then the question of how a repair might affect the \ttt{-UFOO} configuration is itself ill-defined. Input to a repair tool should include at least one macro configuration that is known to compile properly. If a user wants a program to maintain correctness in macro configurations other than the default one specified, the user should indicate which configurations the repair tool must support and preserve.

Most repairs to code will be simple additions of repair macros. For the following analysis, we assume that the repair takes an expression, such as \texttt{a+b},
and replaces it with a macro, such as \verb+acr_safe_add(a, b, goto handle_error)+.

There are several ways that conditional preprocessor directives can interact with this type of repair:

\subsubsection{Independent}

If the expression to be repaired contains no conditional preprocessor directives, we call it \itc{independent}. (The expression may still \itc{be contained between} matching conditional preprocessor directives.)  In this case, the expression is okay to replace with a repaired version; see Fig.~\ref{fig:independent.orig.repaired}.
\begin{figure}[htbp]
\centering
\begin{subfigure}[t]{0.48\linewidth}
\centering
\begin{framed}
\begin{alltt}
  x =
#ifdef WINDOWS
  a+b;  /* expression
to be repaired */
#else /* LINUX */
  c;
#endif
\end{alltt}
\vspace{-2ex}
\end{framed}
\label{fig:independent.orig}
\end{subfigure}
\hfill
\begin{subfigure}[t]{0.48\linewidth}
\centering
\begin{framed}
\begin{alltt}
  x =
#ifdef WINDOWS
  acr_safe_add(a, b,
goto handle_error);
#else /* LINUX */
  c;
#endif
\end{alltt}
\vspace{-2ex}
\end{framed}
\label{fig:independent.repaired}
\end{subfigure}
\vspace{-1ex}
\caption{Independent original (left) and repaired (right).}
\label{fig:independent.orig.repaired}
\vspace{-1ex}
\end{figure}

\subsubsection{Embedded}

For a repair that replaces \texttt{\itc{A}\,+\,\itc{B}} with
\verb+acr_safe_add+\texttt{(\itc{A}, \itc{B}, goto }\verb+handle_error)+,
the \itc{embedded} case is when the subexpressions \itc{A} or \itc{B} contain an \ttt{\#if} conditional preprocessor directive (or a variation such as \ttt{\#ifdef}) and all matching \ttt{\#else}, \ttt{\#elif}, and \ttt{\#endif} conditional preprocessor directives.
Often the embedded case allows the repair to be done safely, as shown in Fig.~\ref{fig:embedded.orig.repaired}.

\begin{figure}[htbp]
\centering
\begin{subfigure}[t]{0.48\linewidth}
\centering
\begin{framed}
\vspace{-0.5ex}
\begin{alltt}
  x =
#ifdef WINDOWS
  a
#else /* LINUX */
  c
#endif
  + b;
\end{alltt}
\end{framed}
\label{fig:embedded.orig}
\end{subfigure}
\hfill
\begin{subfigure}[t]{0.48\linewidth}
\centering
\begin{framed}
\vspace{-1ex}
\begin{alltt}
  x =  acr_safe_add(
#ifdef WINDOWS
  a
#else /* LINUX */
  c
#endif
  , b, goto handle_er
ror);
\end{alltt}
\vspace{-3ex}
\end{framed}
\label{fig:embedded.repaired}
\end{subfigure}
\vspace{-1ex}
\caption{Embedded original (left) and repaired (right).}
\label{fig:embedded.orig.repaired}
\end{figure}
\vspace{-1ex}

However, there are pathological instances of the embedded case, such as in Fig.~\ref{fig:more.difficult.to.repair.example} where \ttt{FOO} is false. Here, repairing the addition if \ttt{FOO} is false will break the code if \ttt{FOO} is true. Likewise, repairing the addition if \ttt{FOO} is true breaks the code if \ttt{FOO} is false. This is because if \ttt{FOO} is true, the conditional contains two statements, but if \ttt{FOO} is false, the conditional contains only one.

\begin{figure}[bht!]
\small
\begin{framed}
\begin{verbatim}
    #include <stdio.h>
    void main() {
        int x=0, y=0, z=0;
        int a=1, b=2, c=3;
        int* e = &c;
        y = (
        #if FOO
        32);
        z = (a
        #else
        * e
        #endif
        ) + c;
        printf("y = %d, z = %d\n", y, z);
    }
\end{verbatim}
\vspace{-2ex}
\end{framed}
\vspace{-2ex}
\caption{Pathological example of conditional directives.}
\label{fig:more.difficult.to.repair.example}
\end{figure}

\subsubsection{Mixed}

In the \itc{Mixed} case, the expression to be repaired contains conditional preprocessor directives in a way that doesn't match the Embedded case. This case is not as easy to repair as the Independent or Embedded cases. In some cases, it is feasible to refactor the code so that it matches the independent or embedded case, as shown in Fig.~\ref{fig:mixed.orig.refactored}. In general, however, the mixed case is not readily repairable.
\begin{figure}[htbp]
\centering
\begin{subfigure}[t]{0.48\linewidth}
\centering
\begin{framed}
\begin{alltt}
x =
#ifdef WINDOWS
  a +
#else /* LINUX */
  a *
#endif
  b;
\end{alltt}
\vspace{-2ex}
\end{framed}
\vspace{-2ex}
\label{fig:original.mixed}
\end{subfigure}
\hfill
\begin{subfigure}[t]{0.48\linewidth}
\centering
\begin{framed}
\begin{alltt}
x =
#ifdef WINDOWS
  a + b
#else /* LINUX */
  a * b
#endif
  ;
\end{alltt}
\vspace{-2ex}
\end{framed}
\vspace{-2ex}
\label{fig:mixed.refactored}
\end{subfigure}
\caption{Mixed original (left) and refactored (right).}
\label{fig:mixed.orig.refactored}
\vspace{-2ex}
\end{figure}

\subsubsection*{Conditional Compilation Directives Outcomes}

We developed a function that takes an expression (specified by its byte range
in the source code) and returns \ttt{True} if it is Independent (i.e., it
contains no conditional preprocessor directives).
If an expression is not Independent, our tool declines to repair it.

\subsection{Trustworthy Repairs}
\label{subsection:trustworthy.repairs}

Making repairs that do not break code can be challenging.
For example, we designed repairs to avoid splitting any expression into multiple statements. This avoids complications that arise when modifying the text of the source code at the character level, rather than at the AST level.

\begin{figure}[bht!]
\small
\begin{framed}
\begin{alltt}
#ifdef RE_ENABLE_I18N
    if (dfa->mb_cur_max > 1)
        bitset_merge (accepts, dfa->sb_char);
    else
#endif
        bitset_set_all (accepts);
\end{alltt}
\vspace{-2ex}
\end{framed}
\vspace{-2ex}
\caption{Excerpt from coreutils.}
\label{fig:excerpt.from.regex.c.coreutils}
\vspace{-1ex}
\end{figure}

For example, consider the snippet from coreutils~\cite{tool.coreutils.regexec.c}
shown in Fig.~\ref{fig:excerpt.from.regex.c.coreutils} (from \ttt{regexec.c}, lines 3549--3554).
Repairing the {\ttt{bitset\_set\_all (accepts)}} line in a way that splits it into two statements would require insertion of curly brackets. For the repair to work with both possible values of \ttt{RE\_ENABLE\_I18N}, the opening curly brace would appear after the \ttt{\#endif}, not on the same line as the \ttt{else} statement.

\begin{figure}[bht!]
\small
\begin{framed}
\begin{alltt}
    for (int x = foo(); x < bar(); x++) \ttc{
        baz();
    }
    int x = 42;
\end{alltt}
\vspace{-2ex}
\end{framed}
\vspace{-2ex}
\caption{A snippet involving a \ttt{for} statement.}
\label{fig:ex.splitting.expressions.into.multiple.requires.relocating}
\vspace{-3ex}
\end{figure}

As another example, consider the snippet in Fig.~\ref{fig:ex.splitting.expressions.into.multiple.requires.relocating}.
Splitting any of the expressions on the \ttt{for} line into multiple statements would require relocating them elsewhere (either inside the body of the \ttt{for} loop or outside the loop).

Splitting these examples is not impossible, but does introduce extra complexity that can be avoided by introducing auxiliary functions so that the expressions can be replaced with other expressions instead of splitting into multiple statements.

\begin{figure}[bht!]
\small
\begin{framed}
\begin{alltt}
trace2/tr2\_tgt\_normal.c:175:49:
        while ((parent\_name = *parent\_names++)) \{
\end{alltt}
\vspace{-2ex}
\end{framed}
\vspace{-2ex}
\caption{Code from Git's \ttt{trace2/tr2\_tgt\_normal.c}.}
\label{fig:code.from.git.tr2_tgt_normal}
\vspace{-3ex}
\end{figure}

Inserting null checks on expressions is more complicated than it seems. Consider the line of code
shown in Fig.~\ref{fig:code.from.git.tr2_tgt_normal}.
Cppcheck warns that \ttt{parent\_names} might be null when incremented and dereferenced.
However, in this code, \ttt{parent\_names} is used as an lvalue~\cite{isoIec98992023}; its value is incremented.
Adding a null check to \ttt{parent\_names} must preserve the lvalue, lest the resulting code fail to compile.
We solve this problem by providing two \ttt{null\_check} macros,
one for rvalues (i.e., expressions that are not used as lvalues)
and one for addressable lvalues (i.e., expressions that represent an assignable memory address). We were unable to find a single macro that solves both cases. We have also not solved the problem for non-addressable lvalues; such a case did not appear in any of the code we tested.

\subsection{Applicability}
\label{subsection:applicability}
We integrated our repair tool into a private Gitlab CI. Our tool's repairs can also be viewed in Visual Studio Code~\cite{tool.ms.vs.code}. Since the repairs each modify only one line of code, they can be easily compared against unrepaired code using a utility such as diff(1); consequently, many tools that can inspect and modify diffs can be used on repaired code. This applies to both Gitlab and VS Code, which provide interfaces for inspecting changes and accepting or rejecting them.

The combination of features in our APR tool may be useful for particular environments.
Its relative simplicity combined with use of Clang AST and LLVM IR might appeal to some APR developers.
It works with the output of 3 SA tools and can easily be extended to others.
It can be used on-premises without requiring expensive high-performance systems, as opposed to some APR tools that require cloud processing or highly performant LLMs, or deep neural networks. 
Although the industry standard for APR use involves patches as suggestions to developers rather than batch processing~\cite{eladawy2024automated}, for some purposes batch processing might be desirable. It can be used on C codebases that don't have test suites yet. Batch repair on a codebase lacking a test suite involves risk if done by our tool, which lacks formal verification of algorithm or software, though other tests of the patched code (compilability, fuzzing, etc. per Section~\ref{section.related.work}) or manual review could provide some degree of confidence in the repairs.

\section{Case Study}
\label{section:case.study}
We implemented repairs for the three flaws and tested them on git and zeek. We conducted several tests to make sure our repairs comply with our constraints.

\subsection{Sample Alert Test}

\begin{table*}[htbp]
  \caption{Alert test results. Each cell has two rows. The first row is the total number of alerts generated; for Git, this cell also contains the number of repaired alerts, the total number of alerts, and the percentage of alerts that were repaired. The second row addresses a sample of $5$ alerts, indicating the number of sample alerts repaired, total number of alerts in the sample, and percentage of alerts repaired satisfactorily.}
\begin{minipage}{\textwidth}
\begin{center}
\begin{tabular}{|c|r|r|r|r|r|r|}
\hline
\bf{} & \bf{git} & \bf{git} & \bf{git} & \bf{zeek} & \bf{zeek} & \bf{zeek} \\
\hline
\bf{Guideline} & \bf{clang-tidy} & \bf{Cppcheck} & \bf{Rosecheckers} & \bf{clang-tidy} & \bf{Cppcheck} & \bf{Rosecheckers} \\
\hline
EXP33-C & 8654 / 9157 (94.5\%) & 1 / 1 (100\%)    & & 5225 &  24 &     \\
        & 5 / 5 (100\%) & 1 / 1 (100\%) & & 5 / 5 (100\%) & 5 / 5 (100\%) & \\
\hline
EXP34-C & 72 / 77 (93.5\%)     & 11 / 20 (55.5\%) & &   44 &  52 &  14 \\
        & 5 / 5 (100\%) & 5 / 5 (100\%) & & 5 / 5 (100\%) & 5 / 5 (100\%) & 5 / 5 (100\%) \\
\hline
MSC12-C &                      & 18 / 25 (72\%)     & &      & 131 & 480 \\
        & & 1 / 5 (20.0\%) & & & 2 / 5 (40.0\%) & \\
\hline
\end{tabular}
\label{table.alerts.count}
\end{center}
\end{minipage}
\vspace{-2ex}
\end{table*}

In this test, we ran our SA tools on our codebases and collected alerts reporting violations of the 3 CERT guidelines we provided repairs for. Table~\ref{table.alerts.count} shows statistics for each combination of tool, codebase, and guideline. Some cells are empty because the SA tool produced no alerts.

The first row indicates the total count of SA alerts. We provide additional analysis for the git alerts, rerunning the SA tools on the repaired version of git, and counting how many alerts remained unrepaired. The first row for each cell contains the number of repaired alerts, total number of alerts, and percentage of repaired alerts to total alerts.

The second row contains sampling data. Since manually validating all of these repairs was impractical given project constraints, we randomly sampled $5$ alerts from each cell ($5$ alerts for each guideline, tool, and codebase), so the results may not achieve statistical significance. We evaluated whether each alert was a true or false positive {\em and} whether our tool {\em should} repair it. Our goal was to improve the tool's reliability at repairing alerts until each cell showed an $80$\% success rate.

 The tool achieved a $100$\% success rate for all EXP33-C and EXP34-C alerts. It correctly repaired or recognized as false all of the null-dereference or uninitialized-variable alerts.

For MSC12-C (code that has no effect), our success rate was lower. MSC12-C alerts were flagged because
\begin{itemize}
\item a local variable was initialized or assigned but never subsequently read;
\item a label was never accessed via \ttt{goto} (This is often code generated by tools like yacc(1).); and/or
\item an unsigned expression was checked for being less than $0$, which is impossible by definition.
\end{itemize}
For these reasons, we did not select $5$ MSC12-C alerts at random, but strove to maximize diversity. We picked $5$ categories of alerts randomly, and then $1$ alert from each category, as a stratified random sample.

Though all of these alerts could be repaired, the repairs would not necessarily improve the code. MSC12-C is a recommendation, not a rule, in the CERT coding standards because it is not always a good idea to make the changes suggested by MSC12-C. Consequently, achieving an $80$\% satisfaction ratio for MSC12-C became a low priority, and these repairs are disabled by default. The failure of the MSC12-C tests to reach the $80$\% repair rate contrasts with the success of the other two repair categories; their successes were not a foregone conclusion. Finally, the MSC12-C repairs are deterministic and can make repairs faster than a human; hence, they are still useful when enabled.

\subsection{Integration Test}

In this test, we built the unrepaired codebase and ran its own testing mechanisms. We then repaired the alerts generated for that codebase. We then rebuilt the codebase and ran it through its own tests. If test results for original and repaired code are identical, our experiment is successful.
Both git and zeek have extensive tests for detecting bugs or regressions. Our repairs had no effect on the output of these tests; therefore, our integration test was successful.

\subsection{Performance Test}

This test confirmed that the repairs imposed on code did not significantly impede performance. In this test, we built and tested the unrepaired codebase and measured the time to completion. We compared this time against the time to build and test the repaired codebase. Ideally the repaired code would be as fast as the unrepaired code.

\begin{table}[htbp]
  \small
  \caption{Timing tests of test suites for Git and Zeek (units \ttt{minutes.seconds} and \ttt{hours:minutes.seconds})}
\vspace{-1ex}
    \begin{center}
      \begin{tabular}{|c|r|r|r|}
        \hline
        \textbf & \textbf{User} & \textbf{System} & \textbf{Elapsed} \\
        \hline
        \textbf{Unrepaired} & 337.99 & 308.69 & 11:24.65 \\
        \textbf{Git} & 341.06 & 318.04 & 11:38.17 \\
        \hline
        \textbf{Repaired}    & 339.08 & 313.82 & 11:31.41 \\
        \textbf{Git}   & 339.54 & 311.17 & 11:28.83 \\
        \hline
        \hline
        \textbf{Unrepaired} & 3.90 & 4.63 & 0:28.43 \\
        \textbf{Zeek}   &  3.50 & 4.71 & 0:28.45 \\
        \hline
        \textbf{Repaired}    & 3.39 & 4.71 & 0:28.54 \\
        \textbf{Zeek} & 3.77 & 4.77 & 0:28.49 \\
        \hline
      \end{tabular}
      \label{table:timing}
    \end{center}
    \vspace{-2ex}
\end{table}

Table~\ref{table:timing} shows the timing data from executions of the test suites for Git and Zeek. Each test suite is invoked by the ``make test'' command. For each test suite, and for each repair state (repaired vs.\ unrepaired), we ran two timing tests.

The average difference in time to run the tests on these repaired and unrepaired codebases was less than the difference between identical runs of these codebases. Performance time was not significantly affected by our code repair.
Two of Zeek's tests timed out after 60 seconds for both the original and repaired versions: \ttt{python-zeek} and \ttt{python-zeek-unsafe-types}.

\subsection{Recurrence Test}

This test confirms that the code was repaired, according to the SA tools. In this test, we repaired the codebases and then ran our SA tools to generate alerts. We compared the alerts from the repaired code to the alerts from the unrepaired code. Ideally, both sets of alerts would be identical except that repaired alerts would be present in the unrepaired set but absent from the repaired set.

\setlength{\tabcolsep}{3.5pt}
\begin{table}[htbp]
  \small
  \caption{Alert counts in recurrence tests for Git and Zeek. (The parenthesized numbers indicate results without the 2 \ttt{sqlite3.c} files in Zeek. See Appendix for details.)}
      \vspace{-1ex}
    \begin{center}
      \begin{tabular}{|c|r|r|r|r|}
        \hline
        \textbf{Guidelines} & \textbf{Clang-tidy}  & \textbf{Clang-tidy} & \textbf{Cppcheck} & \textbf{Cppcheck}\\
        \textbf{Alerts Map to} & \textbf{Original} & \textbf{Repaired} & \textbf{Original} & \textbf{Repaired}\\
        \hline
        \textbf{Git} & & & &\\
        \hline
         EXP33-C &  9157 &   500 &     1 &     0 \\
         EXP34-C &    77 &    16 &    20 &     7 \\
         MSC12-C &     0 &     0 &    25 &     7 \\
        \hline
        All Our 3 & 9234 &   516 &    46 &    14 \\
        \hline
        All Guidelines & 48,233 & 40,840 &   420 &   381 \\
        \hline
        \textbf{Zeek} & & & &\\
        \hline
         EXP33-C &  5225 &   173 &   24 (22) &   13 (3) \\
         EXP34-C &    44 &    14 &   52 (29) &   21 (7) \\
         MSC12-C &     0 &     0 &      &      \\
        \hline
        All Our 3 & 5269 &   187 &   76 (51) &   34 (10) \\
        \hline
        All Guidelines & 18,806 & 15,563 & 2466 (992) & 3071 (899) \\
        \hline
      \end{tabular}
      \label{table:recurrence.git.and.zeek}
    \end{center}
    \vspace{-3ex}
\end{table}
We performed two recurrence tests with Git and Zeek, one with alerts produced by Clang-tidy version 16 and one with alerts produced by Cppcheck 2.9. Table~\ref{table:recurrence.git.and.zeek} summarizes the results. With Clang-tidy, for Git our tool repaired $8718$ alerts,
and Clang-tidy reported $8718$ fewer alerts on the repaired code.
For Zeek with Clang-tidy, all repaired alerts did not recur.
The Cppcheck results are more complicated. For Git, our tool repaired $32$ alerts, but $39$ fewer alerts were reported on the repaired code. So, $7$ alerts disappeared without being explicitly repaired. In Zeek, the number of alerts for our two successful repair categories dropped from $76$ to $34$.
There were problems with repairing and analyzing Zeek with Cppcheck, as detailed in the Appendix.

\section{Related Work}
\label{section.related.work}
The current state of the art and practice in repairing C/C++ code includes much ongoing research, and tools have made significant advances for more than two decades~\cite{legoues-2012-genprog,weimer2010automatic,aberg2003automatic,sidiroglou2005countering,necula2002cil}.
Automated program repair methods commonly use program test failures, SA alerts, and/or bug reports to identify potential defects, localize the code responsible for the defect, produce patches, and run tests to validate correct patching~\cite{eladawy2024automated}. This section discusses some dimensions of APR methods with tool examples (not comprehensive) and comparison to our work.

{\bf{Continuous integration (CI): }} Repairnator
is an open-source modular software repair system for incorporating a wide variety of repair tools into CI systems. The pipeline tries to replicate the bug in a failing CI build and repair it with different APR tools. It can be configured to create pull requests for repairs that pass build tests~\cite{urli2018design}. It could use our APR tool. {\bf{AI: }}Many APR tools use artificial intelligence (AI) to predict code errors~\cite{zhang2023survey}. GitHub's Copilot~\cite{tool.github.copilot} AI tool suggests code, including repairs, and is integrated with widely used integrated development environments (IDEs) and coding tools for many languages, including C/C++. DeepFix is an APR that fixes C language errors by deep learning trained to predict incorrect program locations along with the fixed code~\cite{gupta2017deepfix}. Our tool doesn't use AI.
{\bf{Template-based repairs: }}AVATAR uses fixed patterns of true positives from static analysis to generate patch candidates to fix semantic bugs~\cite{liu2019avatar}. Template-based repairs produce template-based fixes for matching code constructs
and defect categories~\cite{eladawy2024automated}. Our tool makes template-based repairs.

{\bf{IDE integration: }}Widely used IDEs such as Visual Studio Code~\cite{tool.ms.vs.code} and Eclipse~\cite{tool.eclipse} do some automated repairs for C/C++.
The Microsoft Visual Studio IDE provides ``Quick Actions'' that enable applying a code repair for some code flaws, including for SA alerts from external tools~\cite{microsoft.vs.code.quick.actions.cpp}~\cite{microsoft.vs.code.autofixing.and.refactoring.cpp}. In future work, the alert fixes we discuss in this paper could be integrated with that capability. For example, the Snyk plugin~\cite{msvs.snyk.vuln.scanner} provides quick-fix repairs for 15 CWEs~\cite{snyk.vuln.scanner.rules}.
IntRepair detects and analyzes C/C++ flaws and repairs various CWEs as an open-source plugin to the Eclipse IDE~\cite{muntean2019intrepair}.
Its suggested repairs are viewable and can be accepted or rejected in the IDE with a mouse click or done as batch repairs~\cite{tool.intrepair}.

{\bf{Simplicity as repair goal: }}Work by Klieber et al.~\cite{klieber2021automated}
provided automatic rewriting of C code to address memory-safety problems by changing ordinary pointers to ``fat pointers'' that contain the bounds of allocated memory in addition to the raw pointer. Some of these repairs require wide-ranging changes to the codebase, such as when the repair changes the type of an element of a \texttt{struct} or changes the type of an argument to a function with many call-sites. In contrast, our work targets making small local changes, with the goal of fixing many alerts of a few common types of defects. Other tools have similar goals of small repairs; e.g., DirectFix APR tool attempts to find the simplest repairs, using partial MaxSAT constraint solving and component-based program synthesis~\cite{mechtaev2015directfix}. {\bf{Speed and simplicity: }}Unlike some APR tools such as IntRepair, our tool does not use a potentially time-consuming SMT solver to check flaws, instead relying on basic static analysis and compiler tools. Our focus also differs from those auto-repairs to target smaller, simpler-to-fix problems.

{\bf{Preprocessor directives: }}As noted in~\cite{medeiros2017discipline},
many challenges for the current state of automated repair of C/C++ code can be attributed to the C preprocessor, which greatly complicates automated analysis and clean rewrites. That paper advocates a more disciplined approach to using the preprocessor's conditional directives, with preprocessor-oriented rewrites (as opposed to general code repair). Research on rewriting C code recognizes that high-level C or Fortran idioms can be inlined into function calls~\cite{couto2022source}. The development version of the ROS
E compiler framework~\cite{DBLP:conf/jmlc/Schordan-Quinlan-ROSE-2003} supports modifying source code while preserving macros. Software product lines (SPLs) are program families with code mutations (e.g., an \#ifdef directive).
SPLAllRepair is an SPL repair framework that uses variability encoding, bounded model checking, and SAT and SMT solvers on top of the AllRepair APR tool, experimentally resulting in correct C program repairs that were repaired more quickly (linear vs.\ exponential growth) than separately running AllRepair on each configuration~\cite{dimovski2024mutation}. Automated refactoring of preprocessor directives have been applied manually or automatically to improve code understanding and maintainability and enable better use of APR tools~\cite{medeiros2017discipline,van2020tailoring}, e.g., by expanding macros~\cite{reiter2022improving}. Our tool does not provide SPL-level repairs and efficiencies, but it does address some C preprocessor analysis complexities.

{\bf{Binary repairs: }}Mayhem is an autonomous bot that finds and fixes C and C++ vulnerabilities in binaries~\cite{DBLP:journals/ieeesp/AvgerinosBDGNRW18}, while our tool repairs source code.

{\bf{Mutational repairs: }}GenProg uses an extended form of genetic programming to repair code by using existing test suites to encode the defect and its required functionality. Early test results included fixing 8 classes of defects in 1.25M lines of C code in 16 programs~\cite{le2011genprog}.
{\bf{Clang and clang-tidy: }}Clang's~\cite{lattner2008llvm}~\cite{tool.clang} clang-tidy~\cite{tool.clang.tidy} program includes modern refactoring and rewriting APIs~\cite{tool.clang.cre}, and clangd~\cite{tool.clangd} allows for easy integration into text editors and IDEs. Refactoring and rewriting APIs of clangd
are designed to repair only code identified by clang-tidy. Our fixes address some issues that clang-tidy doesn't find (e.g., MSC12-C) and some that clang-tidy detects but does not auto-fix. For example, our tool repairs null pointer dereferences, but clang-tidy doesn't have auto-fixes for its associated checkers (\ttt{core.NonNullParamChecker}, \ttt{core.NullDereference}, \ttt{unix.cstring.NullArg}, \ttt{clang-analyzer-core.NonNullParamChecker}, and \ttt{clang-analyzer-core.NullDereference})~\cite{clang-tidy-checks}.

\def\strutA{\rule[2.75ex]{0pt}{\baselineskip}}
\def\strutB{\rule[2.25ex]{0pt}{0ex}}

\begin{table}[htbp]
  \small
  \caption{Some APR tools that fix the same CWEs in C code as our tool (not comprehensive lists)}
    \begin{center}
      \begin{tabular}{p{0.35\linewidth} | p{0.55\linewidth}}
        \textbf{APR Tool} & \textbf{Notes} \\
        \hline
        \multicolumn{2}{l}{\strutA{}\bf CWE-476 NULL Pointer Dereference} \\
        \hline \strutB
        GenProg~\cite{legoues-2012-genprog}       & Genetic programming \\
        SemFix~\cite{nguyen2013semfix}            & Symbolic execution, constraint solving \\
        Prophet~\cite{long2016automatic}          & Learning-based \\
        SapFix~\cite{marginean2019sapfix}         & Template-based repairs, used at Meta \\
        SPR~\cite{long2015staged}                 & Staged program repair, condition synthesis \\
        FootPatch~\cite{van2018static}            & Repairs SA alerts from Infer w. symbolic heaps\\
        ExtractFix~\cite{gao2021beyond}           & Template-based, constraints from sanitizers\\
        EffFix~\cite{zhang2024efffix}             & Incorrectness separation logic\\
        VulRep~\cite{wei2023vulrep}               & Based on vuln-inducing and fixing commits \\
        VulMatch~\cite{cao5025384enhancing}       & Based on pretrained code model \\
        Flexirepair~\cite{koyuncu2020flexirepair} & Semantic patches \\
        Coccinelle~\cite{lawall2018coccinelle,inria.coccinelle.impact.linux} & Semantic patches, Linux kernel \\
        Conch~\cite{xing2024if}                   & Built on Infer, repairs Infer SA alerts, search- \& constraint-based \\
        \hline
        \multicolumn{2}{l}{\strutA{}\bf CWE-561 Dead Code} \\
        \hline \strutB
        Systematic Code and Asset Removal Framework (SCARF)~\cite{shackleton2023dead} & Used at Meta, human approval if needed \\
        Genetic mutation APR with operators \ttt{delete}, \ttt{insert}, \& \ttt{swap}~\cite{le2012representations} & \ttt{delete} operator removes dead code \\
        \hline
        \multicolumn{2}{l}{\strutA{}\bf CWE-908 Use of Uninitialized Resource} \\
        \hline \strutB
        Prophet~\cite{long2016automatic} & Learning-based, starts with set of good patches \\
        SPR augmented~\cite{mehne2018accelerating}      & Search-based, location-selection, test-case pruning\\
        PAR~\cite{kim2013automatic}      & Manually created repair templates summarize good patches \\
        \hline
      \end{tabular}
      \label{table:apr.tools.three.flaws}
    \end{center}
    \vspace{-3ex}
\end{table}

{\bf{Other APR tools fix the three code flaws in C code that our tool repairs.}}
Table~\ref{table:apr.tools.three.flaws} summarizes some of these tools.

{\bf{Tests: }}For APR tool-use in industry, it is standard practice to run tests after creating a patch candidate (test suites, compilability, general SA tools, dynamic analysis tools such as fuzzers, and sometimes naturalness analyzers), and then for test-passing patches to submit a potential patch for human review before acceptance~\cite{eladawy2024automated,yang2023large,chen2021fast}
Template-based non-learning APR tools like PAR and Redemption don't require the codebase to have a pre-existing test suite. Tools do exist to automatically create test suites, and active research is being done in that area~\cite{lemieux2023codamosa,schafer2023empirical,zhang2011combined,babic2011statically}, but if the new tests fail, work is required to fix the code to make them pass (or to determine the new tests aren't good).
{\bf{Multiple SA tools: }}TRICORDER is a code analysis ecosystem integrated into the production Google development toolchain, including 30 SA tools (as of 2014), and provides APR patches for SA alerts as fix suggestions~\cite{sadowski2015tricorder}. Our tool is not a comprehensive framework but does provide APR fixes for alerts from multiple SA tools.
{\bf{Reduction of SA alerts: }}Google's TRICORDER analysis showed that violations of code flaw checks (checks produce alerts) sharply
change from increasing to decreasing when SA alerts for that flaw appear and APR patches are provided during code \hbox{(self-)review} before check-in~\cite{sadowski2015tricorder}. One striking example of this used patches for clang-tidy checker \ttt{readability-redundant-smartptr-get}~\cite{clang-tidy-checks}.

\section{Lessons Learned}
\label{section.conclusion}
Manually adjudicating and repairing SA alerts is costly, requiring years of work per codebase in our case study.
Our development and test results support the idea that an automatic repair tool could significantly reduce that expense by applying repairs to some of the most frequent alert types.

We learned much while planning and building our repair tool that generalizes for others building APR tools.
We found that some alerts for some secure coding guideline violations are automatically repairable, even if SA tools have poor accuracy in identifying defective code.
This prompted an update of the metrics used to assess when guidelines are detectable or repairable in the CERT Coding Standards.

Many codebases provide their own error-handling routines, while others use standard C constructs like \ttt{abort()}. APR tools should try to use the codebase's error-handling method.

Some C and C++ codebases use conditional preprocessor directives like \ttt{\#ifdef}, which cause a single source file to represent a family of similar source files, not all of which are valid. APR tools should strive not to break the code for any macro configuration or at least to explicitly specify a particular configuration that the repair targets.

We outlined some obstacles with automated repair, such as conditional preprocessor directives appearing inside expressions. We demonstrated in Fig.~\ref{fig:more.difficult.to.repair.example} that it is a simple task to create examples of code that are difficult to repair.

Of the $3$ categories our tool repairs, $2$ satisfied our goal to not break the code, supporting their possible safe batch-repair of multiple alerts. The repair \itc{techniques} do not change the behavior of the code on good execution traces, in contrast to techniques that use genetic algorithms or AI, which cannot make such a guarantee. We do not claim that our tool is free of bugs. Our repair for null-pointer dereferences is to insert code that checks whether the pointer is null immediately before the dereference; thus, our repair does not change the behavior on good traces, where null pointers aren't dereferenced. Our repair for use-before-initialization of variables is to add an initial value at the declaration. Reading an uninitialized variable is undefined behavior in C, so on every good trace, the initial value we assign will be overwritten before the variable is read. Thus, our change of adding an initial value doesn't alter the behavior of any good trace. Our tool's repair for MSC12-C does not change code behavior but is sometimes considered inferior by developers, so it is disabled by default. The repair is deterministic (humans err) and accepting APR patches is faster than human coding, so it can still be useful.

We confirmed that these repairs do not significantly impede performance. Re-running the SA tools on our repaired Git codebase did not produce the repaired alerts or new alerts (except possibly on \ttt{sqlite3.c}; see Appendix).

We learned that many APR tools repair many flaw types with many techniques, some shown to be highly effective against large test sets, some with highly-modular frameworks and scalable test systems, and some widely-used in IDE plugins and industry toolchains.

We published our APR tool, manual adjudications, and auto-code repairs plus the dataset of SA alerts discussed in the paper, so others may build on them or use them to test against other APR tools.
One way to build on our work would be to do similar code flaw prioritization analysis as in this paper, to extend other APR tools to address more code flaw types.
Also, the combination of features in our APR tool may be useful for particular environments.
In the future, our fixes could be incorporated into clang-tidy, using repairs described in this paper for their matching checkers. Our tool could be integrated more deeply into IDEs such as VS Code or Eclipse~\cite{tool.eclipse}. We plan to repair alerts for more coding rules and incorporate techniques for
repairs of multiple macro configurations.

\section*{Acknowledgment}
Copyright 2025 Carnegie Mellon University.
This material is based upon work funded and supported by the United States Department of Defense under Air Force Contract No. FA870225DB003 with Carnegie Mellon University for the operation of the Software Engineering Institute, a federally funded research and development center.
The view, opinions, and/or findings contained in this material are those of the author(s) and should not be construed as an official Government position, policy, or decision, unless designated by other documentation.
NO WARRANTY. THIS CARNEGIE MELLON UNIVERSITY AND SOFTWARE ENGINEERING INSTITUTE MATERIAL IS FURNISHED ON AN "AS-IS" BASIS. CARNEGIE MELLON UNIVERSITY MAKES NO WARRANTIES OF ANY KIND, EITHER EXPRESSED OR IMPLIED, AS TO ANY MATTER INCLUDING, BUT NOT LIMITED TO, WARRANTY OF FITNESS FOR PURPOSE OR MERCHANTABILITY, EXCLUSIVITY, OR RESULTS OBTAINED FROM USE OF THE MATERIAL. CARNEGIE MELLON UNIVERSITY DOES NOT MAKE ANY WARRANTY OF ANY KIND WITH RESPECT TO FREEDOM FROM PATENT, TRADEMARK, OR COPYRIGHT INFRINGEMENT.
[DISTRIBUTION STATEMENT A] This material has been approved for public release and unlimited distribution.  Please see Copyright notice for non-US Government use and distribution.
This work is licensed under a Creative Commons Attribution-NonCommercial 4.0 International License.  Requests for permission for non-licensed uses should be directed to the Software Engineering Institute at permission@sei.cmu.edu.
CERT® is registered in the U.S. Patent and Trademark Office by Carnegie Mellon University.
DM25-1010

\newpage
\bibliographystyle{IEEEtran}
\bibliography{refs.bib}
\section*{Appendix}
\label{section:appendix}
\vspace{-1ex}
We encountered the following problems during the recurrence test for Zeek, with results in Table~\ref{table:recurrence.git.and.zeek}. No other tables were affected by these problems.

First, a different version of Cppcheck was run on the original Zeek than was run on repaired Zeek (2.4.1 vs.\ 2.9). This might be the cause for some variance in the counts.

Second, we had trouble with the two \ttt{sqlite3.c} files embedded in Zeek, for several reasons:
Each sqlite file has about 400 SPLs, and cppcheck takes over 40 hours to analyze one such file. Thus, trying to reproduce previous results on this file was prohibitive.
Also, our tool only repaired one SPL, so it could have no effect on alerts on code that are \ttt{\#ifdef}'d out.
Finally, these files each consist of one C file that contains the entire Sqlite source code. Since Sqlite is external to Zeek, repairs should be made to the sqlite3 source itself rather than to the copy used by Zeek.

Due to time constraints, we were unable to measure the effect of repairing MSC12-C alerts with Zeek.

\end{document}